# WLAN 两种网络架构在地铁 PIDS 系统中适用性的探讨

邓彬

中铁第一勘察设计院集团有限公司 陕西西安 邮编 710043

**摘 要**：本文首先介绍了 WLAN 系统的自治式网络架构与集中式网络架构，比较了两种架构的优缺点，然后详细分析地铁 PIDS 系统对 WLAN 网络的针对性要求，比较 WLAN 两种架构应用于 PIDS 系统的适用性，最后提出地铁 PIDS 系统建设时对 WLAN 网络架构选择的建议。

**关键词**：WLAN 自治式架构 集中式架构 地铁 PIDS

## 引言

地铁内乘客信息显示系统（PIDS：Passenger Information Display System）主要承载两类业务。第一，为乘客搭乘地铁全程提供咨询服务（包括候车与乘车），如：发布交通信息、公益广告、运营紧急救灾信息等，此类多媒体信息由地铁 PIDS 编播中心向各车站及运营中的列车下发。第二，PIDS 系统承担将运营中的列车各节车厢内的视频监控图像上传至相应控制中心的任务。

运行中的列车与编播中心或控制中心交换上述信息时，需要借助 PIDS 中的车地无线子系统完成双向无线宽带通信功能。目前，国内地铁多采用无线局域网（WLAN：Wireless LAN）来构成车地无线子系统。

## 1 简介

采用符合 802.11 系列无线局域网标准的 WLAN 构成的车地无线子系统，主要设备有无线接入点 AP（Access Point）、无线接入点控制器 AC（AP Controller）或 WLAN 交换机、车载无线客户端网桥、隧道天线等，配合 PIDS 系统中的车站级交换机、核心交换机、中心网络设备、信源、车载交换机、车载显示系统、有线传输通道等，组成完整的系统。PIDS 系统网络结构示意图如下：

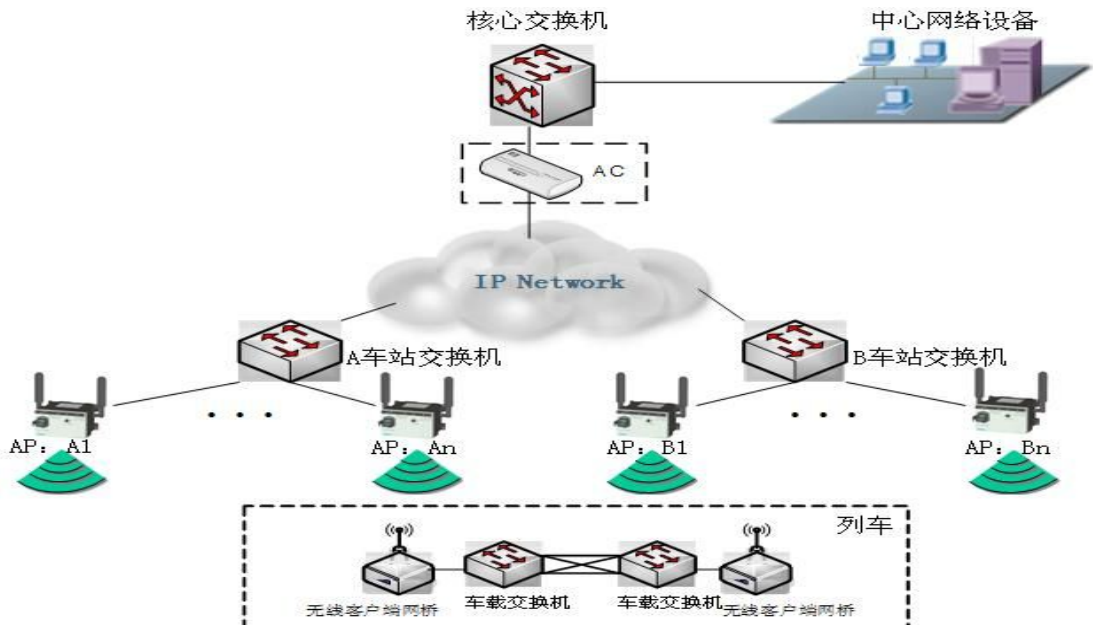

图 1 PIDS 系统网络结构示意图（本图为集中式架构）

## 2 自治式架构与集中式架构[1]

WLAN 的网络架构主要有自治式架构与集中式架构两种。

自治式架构中，AP 独自完成 802.11 标准中规定的所有功能。每个 AP 均是网络上的一个单独的网络实体，自身独立进行管理。

集中式架构中，网络中所有 AP 由一个或多个 AC 负责配置、控制和管理，802.11 功能由 AP 和 AC 共同分担。与自治式架构相比，这种模式中的 AP 一般只负责完成无线通信的物理层的相关功能。

两种架构的本质区别是将 802.11 中的网络控制和管理功能采用分布式结构还是集中式结构来实现，不同的实现方式各具特点，下面对两种架构进行比较分析。

### 2.1 自治式架构

在 WLAN 部署的初期，无线局域网主要定位于满足办公场所或家庭用户的互联网无线接入，一般在需要的场所直接设置 AP 提供无线覆盖。在此阶段，几乎所有的 AP 都是自治式 AP，独立完成用户终端设备接入无线网络的加密、终端设备管理、无线信号收发、帧格式转换等功能。由于这种 AP 设备承担多种网络功能，通常被称为"胖 AP（Fat AP）"。

"胖 AP"独立完成自身管理，不对网络中高层汇聚设备返回信息。一般需要网管人员逐个进行参数配置，如：使用第几号子信道、最大发射功率、SSID、加密方式以及一些路由和安全策略。

自治式架构的结构示意图见图 2 所示。

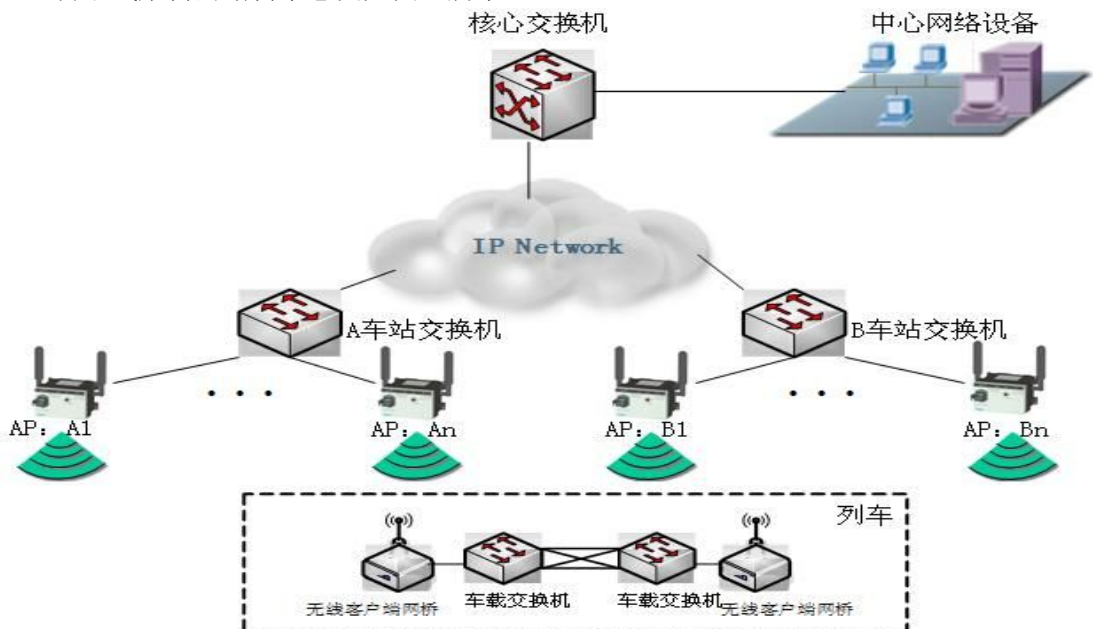

图 2 自治式架构的结构示意图

### 2.2 集中式架构

随着无线局域网越来越广泛的使用，企业或政府部门开始部署数量庞大的 AP 用来覆盖更多的地域。要完全覆盖一个大型企业一般需要几百个至上千个 AP，而"无线城市"至少需要上万个 AP。对于数量如此众多且密集分布的 AP 布置，传统的"胖 AP"逐渐暴露出一些弊端。例如：密集分布的 AP 之间难以避免信道间的干扰、每个 AP 的业务负荷很不均匀、网络建成后的管理难度大等。这些问题，只有少部分可以在网络建设初期通过合理规划、优化设备参数配置等方式解决或降低影响，大部分问题都是传统的"胖 AP"无法克服的。

以上问题，究其根源是因为 AP 各自为政，网络对 AP 缺乏统一有效的协调和管理引起的。为此，在网络中增加新的网元—AC 设备，用来集中完成无线网络管理的相关功能，而将原有 AP 的功能弱化，令其只负责无线信息收发，这种功能相对单一的 AP 被称为"瘦 AP（Fit AP）"。"瘦 AP"需要向 AC 报告与网络管理有关的信息，两者之间一般采用 CAPWAP 协议[2]。

集中式架构将网络管理功能集中在 AC 设备上，AC 可以直接收集由 AP 或相关硬件设备感知的网络前端的各种统计信息，具有透视整个下层网络的能力。

集中式架构的结构示意图见图 1 所示。

## 3 两种架构的比较

自治式架构与集中式架构各具特点，下面简单比较各自的优缺点[3]：

从网络设备组成上比较，自治式架构只需要 AP，设备种类少；集中式架构需要增加 AC，AP 需要在 AC 处做一次汇聚。

从网络管理能力上比较，自治式架构网管功能分解于每一个 AP 自身，管理功能受限，难以支持整网管理策略；集中式架构在 AC 处进行统一管理，可按需定制及调整管理策略，方便增加管理功能，管理能力强。

从网络可靠性上比较，自治式架构中，个别 AP 故障不影响全网功能；集中式架构中，若 AC 故障将引起全网瘫痪。

从切换时延上比较，自治式架构中，用户终端在 AP 之间漫游时，时延在几百毫秒不等；集中式架构中，用户终端在同一个 AC 管理下的 AP 之间漫游时，时延在数百毫秒。

从网络扩展能力上比较，自治式架构网络中 AP 数量越多，弊端就越严重；集中式架构可通过升级至更大容量的 AC，或通过增加 AC 数量的方式解决扩容的问题。

从增加新功能上比较，自治式架构的网络一方面无法具备需要全网信息作为输入的新功能，另一方面增加新功能需要涉及到每一个 AP 的升级；集中式架构中，AC 具备全网观测的能力，增加新功能一般只需升级 AC 即可完成。

从安全性上比较，自治式架构的安全策略在每个 AP 上独立配置完成，一般只有基本的安全能力；集中式架构可以在 AC 上配置全网统一的安全策略，一般有较完善的入侵检测与防御能力，并可以开发专有的安全机制。

从网络负载均衡上比较，自治式架构的 AP 无法协调网络负载，有可能出现一个 AP 超负荷而其邻近 AP 又空闲的情况；集中式架构中，有些 AC 具有动态协调 AP 负载的功能，使网络不易发生阻塞。

两种架构的优缺点比较见表 1：

表 1 两种架构的优缺点比较表

| | 自治式架构 | 集中式架构 |
| --- | --- | --- |
| 网络设备组成 | 少 | 多 |
| 网络管理能力 | 弱 | 强 |
| 网络可靠性 | 高 | 存在单点故障 |
| 切换时延 | 较大 | 较大 |
| 网络扩展能力 | 弱 | 强 |
| 网络增加新功能 | 困难 | 容易 |
| 网络安全 | 一般 | 强 |
| 网络负载均衡 | 无均衡能力 | 有均衡能力 |

## 4 两种架构在地铁 PIDS 系统中的适应性分析

地铁 PIDS 系统为专网系统，与公网系统相比有自己的特殊需求与侧重点，以下针对地铁的应用场景，分析两种架构在地铁 PIDS 系统中的适用性。

（1）设备可靠性及防护等级

地铁中采用的机电设备，首先必须可靠耐用。地铁隧道内潮湿、多尘、震动大，高架区间雨淋、日晒，现场环境恶劣。这要求安装于轨旁的 AP 必须是户外型设备，并且要具有不低于 IP65 的防护等级。目前，两种架构的 AP 设备均有满足地铁应用环境要求的成熟产品，两种架构均可采用。

（2）网络可靠性

地铁 PIDS 系统要求网络中单个设备的故障不能影响全网的功能。自治式架构中，AP 的单点故障不影响网络。集中式架构中由于 AC 的汇聚作用，当 AC 故障时将导致全网瘫痪。为解决这一问题，可以增加一套 AC 进行热备份。具有 AC 热备的集中式架构与自治式架构的网络可靠性相差无几。

（3）切换时延

当列车运行时，列车上的无线网络终端不可避免的要在AP之间发生切换（或称为漫游）。切换过程会对信息的传输带来时延，对于以视频为代表的实时型业务（如车内的监控视频上传、车站视频广告下发至列车显示系统等），过大的时延将会影响视频的顺畅播放，严重的时延将会导致视频的中断。

两种网络架构中，终端均以硬切换的方式完成切换功能，即终端首先与正在连接的 AP 设备断开连接，再与新的 AP 设备发起关联。切换基本可分为三个阶段进行，即发现新 AP 阶段、认证阶段和重连阶段。有资料显示，在基本忽略认证阶段的情况下，切换时延大致在数百毫秒，且不同厂家的设备差异显著[4]。实际中如果网络认证算法复杂，切换时延还将增长[5]。目前，关于提高切换性能的研究很活跃，有多种新算法被提出，均称能够大幅减少切换时延[6][7]。但是由于测试的平台和标准不统一，难以有一个公平可信的数据。

值得注意的是，目前一些厂商在产品中可能集成一些非 802.11 标准的切换协议，用来更好的支持实时业务的切换。如西门子的具有快速漫游功能（iPCF）的产品[8]及中兴的可管理快速切换技术(MFHO)[9]等，均称可以将切换时延控制在 50ms 左右。但相关资料中缺少详细的测试网络软硬件及采用协议的相关信息。

（4）建设与维护的便利

PIDS 系统的 AP 在单隧道内一般每隔 100 至 200 米布置一套，一条地铁线路一般需要数百套 AP。AP 的参数配置及后期维护的便利是工程设计中必须考虑的问题。

自治式架构中，AP 的参数需要逐个进行配置。虽然某些产品具有利用 U 盘存储参数简化配置的功能，但仍然需要网管人员在每个 AP 处现场操作。人工配置参数的工作量大，容易出错。网络建成后的运营维护，如果需要调整网络参数，也必须逐个 AP 进行参数修改。

集中式架构中，无线网络的各设备参数及管理策略集中在 AC 处配置，AP 与 AC 间采用 CAPWAP 协议协商管理。AP 在第一次接入网络时会自动与 AC 进行通信，获取相关参数配置。运营维护期间，如需要改变网络设置，只需在 AC 处做相关升级，无线网络的参数即可做相应更新。

集中式架构在管理的自动化及先进性上具有自治式架构无法比拟的优越性。

（5）网络安全

地铁 PIDS 系统主要传输为旅客服务的信息，即使传输的信息被窃取也不存在安全问题。为防范无意接入或一般入侵，不论是自治式还是集中式架构中的常规安全功能，如口令及密钥等，均能够提供足够的安全能力。

但是地铁 PIDS 系统作为一个影响公众人数众多的信息发布平台，需要考虑防范恶意入

侵者发布非法视频信息等内容。此外，某些城市的地铁考虑利用 PIDS 系统承载更多高等级的车厢内安防功能，如车厢内生化攻击的实时监控与检测等。预计这些业务将对无线网络提出比目前更高的安全要求。

自治式架构的网络在面对伪装接入点等网络欺骗手段时，需要很高的代价才可能堵住漏洞；面对日新月异的网络攻击手段，其安全策略不够灵活，难以进行快速的有针对性的调整。

集中式网络架构中的 AC 设备，结合无线资源管理功能，可以快速发现伪装的接入点，并且可以命令邻近的合法 AP 提高同一个信道上的发射功率来屏蔽这个伪装的接入点。对于不断更新的各种攻击手段，有针对性的全网安全策略可以经由 AC 快速部署。更重要的是，AC 的全网透视能力可以综合分析更高层面的网络运行数据，及时发现潜在的风险。

（6）干扰

隧道区间内的 PIDS 系统，由于空间封闭且深埋地下，与外界各种随机干扰源有足够的隔离，因此隧道内的无线环境比较单纯。在设计阶段为不同的无线系统规划好相应频率，就可以将干扰控制在可接收的范围内。

地铁高架区间的 PIDS 系统，会受到地铁外的各种随机干扰源的干扰，并且无法提前预测。此时，自治式架构的受干扰的 AP 吞吐能力会降低，严重时无法正常工作。而集中式架构中，AC 具有无线资源管理功能（需配合具有无线环境监测功能的 AP 或其他硬件设备），当干扰发生时会命令受干扰的 AP 选取其他无线信道，以规避干扰。无线信道检测及调整的过程自动完成，大大降低干扰的危害。

（7）设备故障

PIDS 系统中 AP 的布置需要满足冗余覆盖的要求，以保证当网络中某个 AP 故障时，邻近的两个 AP 还可以提供无盲区的覆盖，减小终端脱网的概率。

在自治式架构的网络中，首先需要根据 AP 最大发射功率规划每个 AP 的覆盖范围，达到冗余的要求。规划的结果使得网络在正常运行时，区间的任何地点都同时存在至少两个 AP 的信号，这导致区间的无线环境比较复杂，容易发生乒乓切换等意想不到的情况。另外，规划完毕后，当网络中由于其他原因需要再增加 AP 时，必须重新设定相邻多个 AP 的无线发射功率等参数。

在集中式架构的网络中，AC 可以根据每个 AP 的覆盖情况，自主协调控制 AP 的信号功率。在网络正常时，区间任何地点都只有一个 AP 的信号（除切换覆盖区）；当有 AP 出现故障时，AC 会自动检测到故障，并指示邻近的 AP 调整功率和信道设置来弥补发生故障的 AP 留下的盲区。当新的 AP 加入网络时，AC 也会命令其他 AP 减小功率发射，以避免产生干扰。这种功率自适应的功能，对工程设计和运营维护都很有利，使网络可以智能的适应实际环境的变化。AP 覆盖的示意图见图 3：

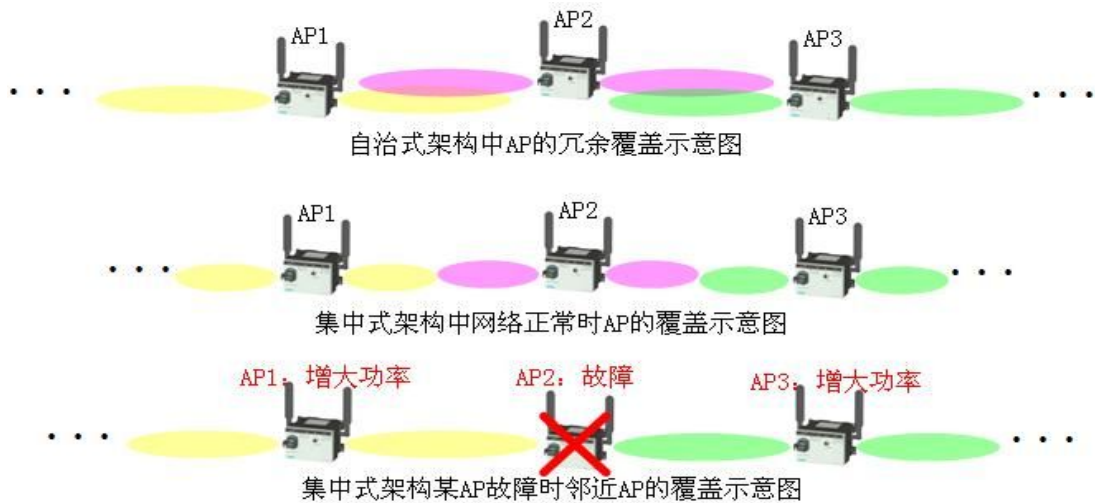

自治式架构中AP的冗余覆盖示意图

集中式架构中网络正常时AP的覆盖示意图

集中式架构某AP故障时邻近AP的覆盖示意图

图 3 AP 覆盖示意图

## 5 建议

以上分析了两种架构在 PIDS 系统中的适应性。笔者认为，设备可靠性、网络可靠性与切换时延直接关系到 PIDS 系统是否能够正常使用，是最基本的技术指标，在选择如何构建网络中应具有最大的权重。前两项参数不论自治式或集中式架构均可以很好的满足。切换时延比较复杂，建议有条件时将各投标厂家设备搭建试验网络进行实测，无条件时请投标厂家出具切换时延的保证函，供招标选择。

其他比较的参数和项目对 PIDS 系统进行软支持，如果考虑到网络进一步的发展，集中式架构相比能够提供更好的扩展能力。